\documentclass[12pt]{nature}
\usepackage[square, numbers, comma, sort&compress]{natbib}

\usepackage{color, times}
\usepackage{graphicx, amsmath, amssymb, caption}
\usepackage{placeins}
\usepackage{supertabular}
\usepackage{xspace}
\usepackage{longtable}
\usepackage{dcolumn}%Align table columns on decimal point
\usepackage{bm}%bold math
\usepackage{physics}
\usepackage{xfrac}

\usepackage[mathscr,scaled=1.15]{urwchancal}
\DeclareFontFamily{OT1}{pzc}{}
\DeclareFontShape{OT1}{pzc}{m}{it}%
{<-> s * [1.15] pzcmi7t}{}
\DeclareMathAlphabet{\mathpzc}{OT1}{pzc}{m}{it}

\definecolor{purple}{rgb}{0.5,0,0.5}
\definecolor{blue}{rgb}{0.0,0,0.9}
\definecolor{prdblue}{rgb}{0.133,0.118,0.498}
\usepackage[colorlinks=true, pdfstartview=FitV, linkcolor=prdblue, citecolor= prdblue, urlcolor=prdblue]{hyperref}

\title{$\,$\\[-4ex]{\normalsize Artificial Dynamical Effects in Quantum Field Theory}}
\author
{Stanley J. Brodsky,$^{1,\ast}$ Alexandre Deur,$^{2}$ Craig D. Roberts$^{3,4}$\\
\\
\normalsize{$^{1}$SLAC National Accelerator Laboratory, Stanford University,
Stanford, California 94309, USA}\\
\normalsize{$^{2}$ Thomas Jefferson National Accelerator Facility, Newport News, VA 23606, USA}\\
\normalsize{$^{3}$ School of Physics, Nanjing University, Nanjing, Jiangsu 210093, China}\\
\normalsize{$^{4}$ Institute for Nonperturbative Physics, Nanjing University, Nanjing, Jiangsu 210093, China}\\
{\normalsize
E-mail: sjbth@slac.stanford.edu; deurpam@jlab.org; cdroberts@nju.edu.cn}\\
{\normalsize $^\ast$Corresponding author}
}

\date{2021 July 20}
\begin{document}

%Double-space the manuscript.

\baselineskip24pt

\maketitle

\begin{abstract}
In Newtonian mechanics, inertial pseudoforces -- or fictitious forces -- appear in systems studied in non-Galilean reference frames; e.g., a centrifugal force seems to %act on objects
arise
if the dynamics is analyzed in a rotating reference frame.
The equivalent of Galilean invariance for relativistic kinematics is Poincar\'e invariance;  analogous artificial effects may arise in relativistic quantum field theory (QFT) if a system is
studied in a framework violating Poincar\'e invariance.
We highlight how such issues complicate the traditional canonical quantization of QFTs and can lead to a subjective description of natural phenomena.
In fact, if the system involves the strong interaction, obtaining %Poincar\'e invariant 
objective results can become an intractable problem using canonical quantization because the pseudoforces are essentially nonperturbative.
In contrast, the treatment of the same problem using light-front (LF) quantization is free of spurious pseudoeffects because Poincar\'e invariance is manifest; thus the treatment of strong interaction problems becomes simpler.
These statements are illustrated using several examples:
the Gerasimov-Drell-Hearn (GDH) relation, a fundamental feature of QFT;
the absence of any measurable impact of Lorentz contraction in high-energy collisions;
and the fictitious character of vacuum fluctuation contributions to the cosmological constant.
\end{abstract}

\section{Introduction }
\label{introduction}
A key scientific principle is that an observer's choice of conventions cannot affect physical phenomena. Effects that depend on such choices are artificial, rather than objective features of Nature; e.g., physics cannot depend on the choice of units or a particular scheme to renormalize a quantum field theory.  Most critically, physical effects cannot depend on the observer's Lorentz frame: valid phenomena should display Galilean invariance in classical mechanics and Poincar\'e invariance for relativistic systems.

In Newtonian dynamics, inertial (Galilean) frames are essential, because they provide a minimal description of Nature: only fundamental forces appear. In non-Galilean frames, artificial inertial forces can arise, e.g., centrifugal forces which seem to act on an object from the perspective of an observer in a rotating frame. Such \emph{ad-hoc} frame-dependent pseudoforces are not innate to the dynamics of the system. Although non-Galilean frames and pseudoforces are sometimes  intuitively useful, the fundamental description of a system in Newtonian dynamics must be determined in a Galilean frame.

Just as breaking Galilean invariance in Newtonian dynamics introduces artificial pseudoforces, any approach to relativistic QFT which breaks Poincar\'e invariance may also lead to pseudophysics. In order to identify situations with such potential, recall the two common methods of field quantization: the Hamiltonian-based canonical approach; and Lagrangian path integrals.

\noindent{\bf Canonical}.
This approach uses the Hamiltonian $\mathcal H= i{d\over {d{\mathpzc t}}}$ as the generator of evolution in time ${\mathpzc t}$.
Since it involves time-ordering, it is also called ``time-ordered perturbation theory''.
However, by singling-out the time coordinate $t$, which is a non-Lorentz-invariant variable, one breaks Poincar\'e invariance; thus potentially introducing artificial pseudodynamical effects, as discussed below.

\noindent{\bf Path integral}.
This Lagrangian approach leads to ``covariant perturbation theory'', centered on Feynman diagrams;  time-ordering is not required.
Thus, for a reaction involving $n$ vertices, a single Feynman diagram corresponds to the sum of $n!$ time-ordered diagrams.
Since time or coordinate frame definitions are not needed, covariant perturbation theory
is free of pseudodynamics.
It is natural to make frame and metric choices to produce a particular prediction from the covariant calculation.
This is also standard in Newtonian dynamics.
For example, to connect with a practical situation, one may choose a specific frame and spacetime metric, such as the Minkowski metric, $(t,x,y,z)$, or the LF metric~\cite{Dirac:1949cp},  $(x^-,x,y,x^+)$, $x^\pm = ct \pm z$. 
%(Henceforth, natural units are used, e.g., $\hbar \equiv 1 \equiv c$.)
(Henceforth, we set $\hbar \equiv 1 \equiv c$.)

Although the canonical and path integral quantizations can be used independently, they are often combined. To avoid the consequences of breaking Poincar\'e invariance in the canonical approach, a perturbative calculation is typically developed by employing the Hamiltonian to generate the dynamics, summing over all time-orderings, and then deriving the Feynman rules for use in calculations, thereby guaranteeing covariant results. Using solely the Hamiltonian approach may seem a poor choice, since it exposes practitioners to the issues of Poincar\'e invariance violation, in contrast to covariant perturbation theory or the Hamiltonian approach supplemented by Feynman graphs. However, physical processes may also involve nonperturbative dynamics, which is not calculable using Feynman diagrams at any finite order.

Nonperturbative dynamics can emerge from Feynman graphs when infinitely many diagrams are resummed, as expressed, e.g., in the Dyson-Schwinger equations (DSEs) \cite{Roberts:1994dr}. The perspective can be reversed, with nonperturbative DSEs serving as generating tools for Feynman graphs. However, applying DSEs %to physics problems requires truncations; and developing sound, symmetry preserving truncations is complex \cite{Eichmann:2016yit, Qin:2020rad, Qin:2020jig, Roberts:2021nhw}. %\cite{Qin:2020rad, Qin:2020jig, Eichmann:2016yit, Burkert:2017djo, Fischer:2018sdj, Dupuis:2020fhh, Barabanov:2020jvn, Roberts:2021xnz}
involves the development of sound, symmetry preserving truncations, which is a complex task \cite{Eichmann:2016yit, Qin:2020rad, Qin:2020jig, Roberts:2021nhw}.

The Hamiltonian approach also offers a natural framework to formalize nonperturbative dynamics, albeit with the caveat that pseudoforces may arise.
Herein, we concern ourselves with this method and how to avoid pseudodynamics \cite{Dirac:1949cp, Brodsky:1997de}, offering new perspectives by drawing parallels with well-known Newtonian dynamics. This discussion will aid in identifying the origin of the caveats that ride on the Hamiltonian approach and how to avoid them.

%%%%%%%%%%%%%%%%%%%%%%%%%%%%%%%%%%%%%%%%%%%%%%%%%%
\section{Source of Poincar\'e invariance violation}
\label{Poincare invariance violation}
Canonical quantization is typically completed in Minkowski spacetime and called ``instant form quantization'' (IF) \cite{Dirac:1949cp}.
IF is widely used because of its links to nonrelativistic systems and ensuing intuitive nature.
However, applying IF quantization to systems with relativistic dynamics is problematic because it manifestly violates Poincar\'e invariance.
LF quantization, the other common canonical procedure, 
%is largely immune to this 
avoids the issue since it retains boost invariance and is independent of the observer's Lorentz frame. 
%%%AD: I am commenting the following because we mention it later, and here seems not to best place.%%%
%Observations in LF time $\tau= x^+ = ct +z$ are analogous to flash photogaphs.

Both forms of relativistic dynamics \cite{Dirac:1949cp} are initially constructed in compliance with Poincar\'e invariance.
In both cases, one identifies 10 operators which fulfill the Poincar\'e algebra, expressing the spacetime symmetries, {\it viz}.\ invariances under spacetime translations, spatial rotations, and boosts.
However, Poincar\'e invariance concerns the worldlines of four-dimensional (4D) spacetime.
To study the time evolution of a system in 3D space, spacetime must be foliated to specify the time variable.
The choice of foliation is the difference between IF and LF. %, or other canonical quantization forms.

In general, a symmetry manifest in 4D need not be preserved in a subspace of lower dimension; and in 3D, no canonical quantization form will satisfy all the requirements of Poincar\'e invariance.
The following analogy reveals how Poincar\'e invariance becomes lost in IF, but is preserved in practice by LF.
In 3D, a cylinder is rotationally invariant around one axis, conventionally labelled $\bm z$. %$\hat z$. %%AD: replaced \hat by \bm for style consistency with the rest of the article
If the cylinder is projected onto a plane, then the symmetry can either be maintained or broken.
When %$\hat z$ 
$\bm z$ is perpendicular to the plane, the projection is a circle and the symmetry is preserved.
Otherwise, the projection is an ellipse and rotational invariance is lost.
In this analogy, LF corresponds to a ``perpendicular'' slicing that preserves the symmetries of primary interest; whereas IF corresponds to an ``elliptical'' slicing, %for which the projection leads to violation of a desired symmetry (frame invariance); 
that breaks the desired symmetries; 
a crucial problem-solving asset is thus lost.
Notably, LF retains the maximum number of {\it kinematical} operators: 7 out of 10%. These operators leave invariant the initial hypersurface, {\it viz}.\ the 3D space defined by a fixed initial light-front (LF) time, thereby preserving Poincar\'e symmetry.
, that do not involve dynamics (genuine or pseudo).

Crucially, the three {\it dynamical} LF operators %, those that change the initial hypersurface, 
are the LF Hamiltonian, which describes evolution in LF time $x^+ = ct +z$, and two rotation operators in %$\hat x$ and $\hat y$
$\bm x$ and $\bm y$ which are often unimportant for reactions in high-energy physics.
Consequently, for practical purposes, LF dynamics is effectively Poincar\'e invariant.
In contrast, IF has only 6 kinematical operators (the operators for space translations and rotations). Its boost operators are dynamical: they introduce interactions in Lorentz transformations,  %in the transformations that correspond to important symmetries in the 3D space,  
complicating the IF treatment of reactions.
Thus, Poincar\'e invariance can in practice be violated;
%Indeed, IF Lorentz boosts necessarily mix kinematical and dynamical dependence. Therefore, 
and a phenomenon observed in two distinct frames related by different boosts can have contrasting dynamical descriptions.
%%%AD: Below is a repeat of the previous statement. %%%
%Calculations in the distinct frames must produce the same result, but details concerning which forces are relevant in the IF will depend on the boost, a manifestation of the volation of Poincar\'e invariance.
This introduces severe problems: the dynamical description becomes subjective, depending on the observer's motion. Furthermore, it may not be possible in practice to identify and compute the boost-dependent pseudoforce.

\begin{figure}[t]
\centering
\vspace{-0.cm}
\captionsetup{width=0.88\linewidth}
\includegraphics[width=6.0cm]{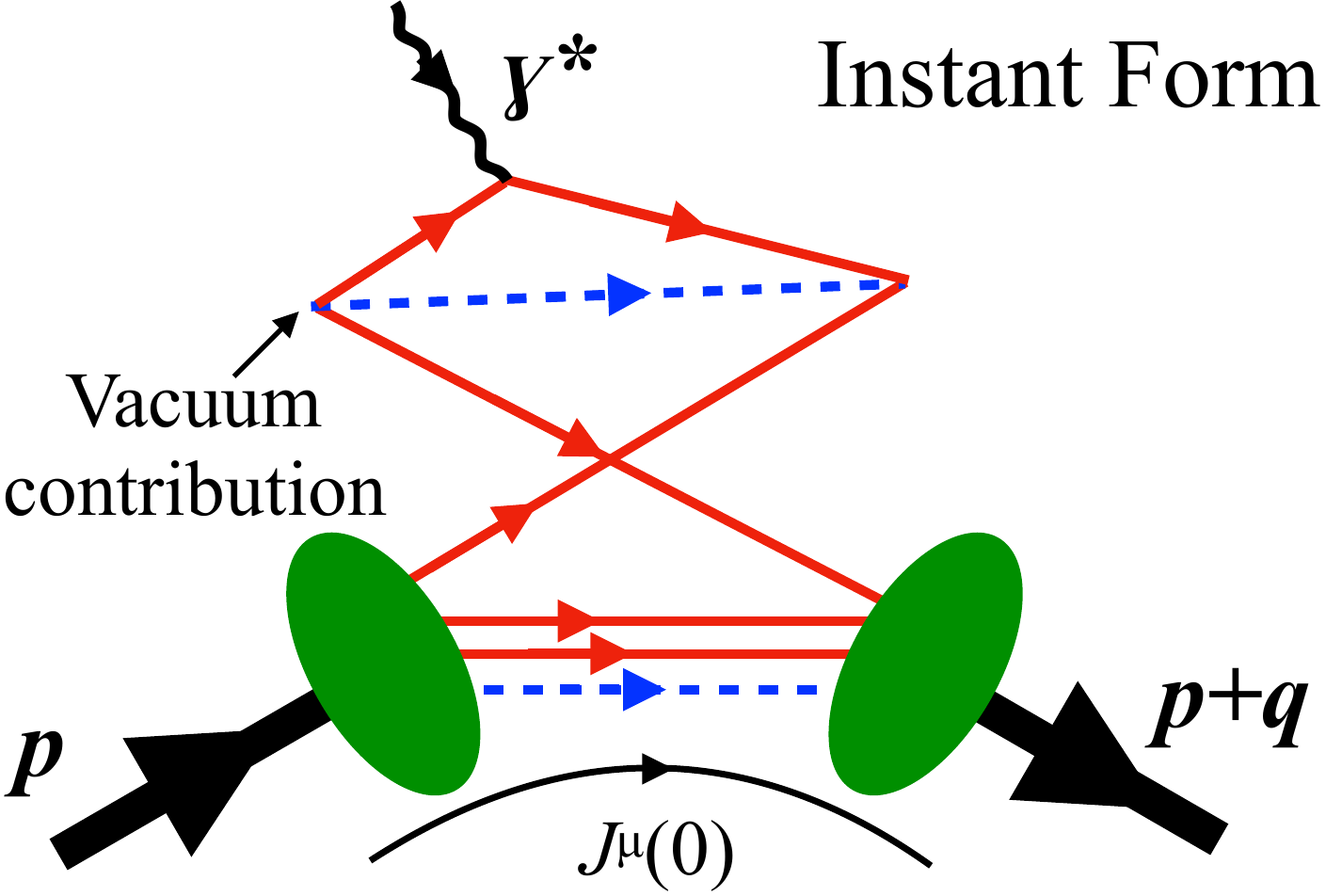}
\hspace{1.cm}
\includegraphics[width=6.0cm]{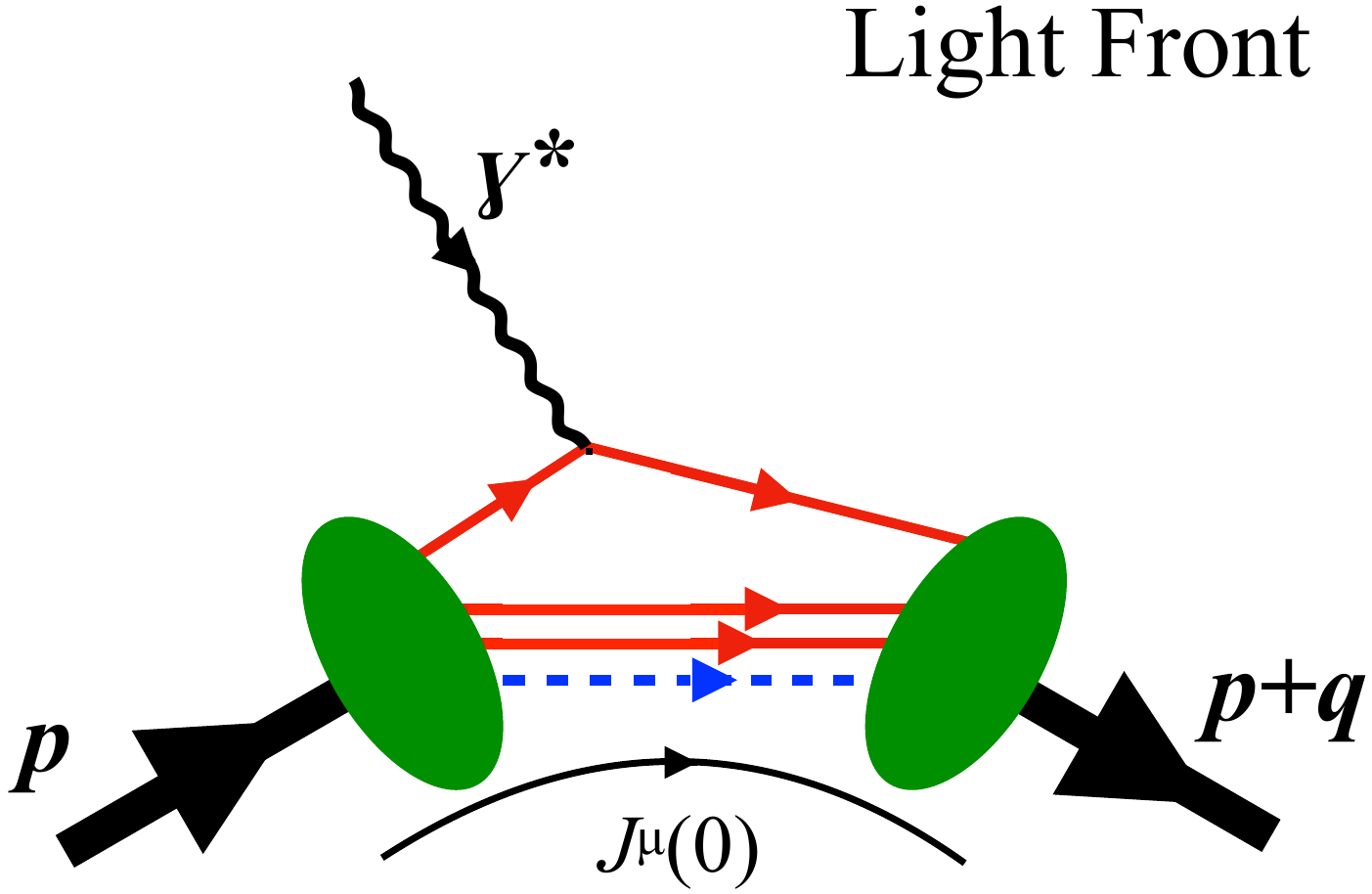}
%\vspace{-0.3cm}
\caption{\label{Flo:Form-Factor in the IF}
%{\footnotesize }
{\bf Graphs used to compute the nucleon electromagnetic form factors in the IF (left) and LF (right) cases.} A virtual photon $\gamma^*$ (wavy line) probes a nucleon (thick black line) with initial momentum $p$ and final momentum $p+q$.
The electromagnetic current at spacetime location $x$ is $J^\mu (x)$.
The red solid and blue dashed lines denote valence quarks and gluons, respectively.
Acausal, frame-dependent vacuum-induced currents are required in IF calculations of current matrix elements -- left.  
They are absent in LF -- right.
}
\end{figure}

The mixing between dynamical and kinematical effects makes the IF boost of a bound state wave function a complex dynamical problem which may not be resolved within the IF framework alone \cite{Brodsky:1968xc, Brodsky:1968ea}, as highlighted in Sec.\,\ref{SecGDH}.
The fact that IF boosts are dynamical is a major hurdle because boosts are essential in analyzing high-energy reactions, e.g., elastic electron+proton scattering,\linebreak $e+P(p) \to e^\prime+P^\prime(p+q)$, where the electron transfers a 4-momentum $q$ to a proton of initial 4-momentum $p$, as illustrated in Fig.~\ref{Flo:Form-Factor in the IF}.
To study such reactions, the proton wave function must be boosted from $p \to p+q$.
For IF, this leads to a complicated dynamical problem, compounded by the fact that the number of particles describing the proton can change, as will be discussed next.
The boost of a bound state combines nonperturbative dynamics with kinematics, typically making
the IF boost of a bound-state wave function in quantum chromodynamics (QCD) incalculable, and thus the analytic determination of the boost-dependent pseudoforce intractable.

%In addition to the appearance of pseudoforces, there
%
There is another complication when using IF: it cannot describe a system's dynamics while respecting causality.
This has far-reaching consequences.
For instance, the form factor (Fig.~\ref{Flo:Form-Factor in the IF}) of a bound state involves current matrix elements, like $\bra {p+q}J^\mu(0) \ket {p} $, which cannot be computed solely from the bound state IF wave function because additional currents arise from pair creation from the vacuum.
For example for the muonium atom ($\mu^+ e^-$), one must also couple the external current $J^\mu$ to the leptons that are created or annihilated from the vacuum by the $e^+ e^- \gamma $ interaction.
These ``vacuum-induced" effects appear when one relates Feynman diagrams to the sum of canonical IF perturbation theory diagrams, as illustrated in Fig.~\ref{Flo:Form-Factor in the IF}.
Such contributions are acausal because the pair-creation time is uncorrelated with the time of the scattering event.
Thus, Poincar\'e-invariant physical results cannot be obtained using the IF without including acausal dynamical contributions.
In contrast, pair contributions are absent in the LF framework, and form factors %of bound states 
can be correctly computed \cite{Drell:1969km, West:1970av, Brodsky:1980zm}. %We will now summarize some of its key features.

%%%%%%%%%%%%%%%%%%%%%%%%%%%%%%%%%%%%%%%

\section{The Light Front \label{LF basics}}
The fundamental problems arising from IF vanish with LF quantization \cite{Dirac:1949cp, Brodsky:1997de,Mannheim:2020rod}, which is effectively frame-independent and causal.
The Galilean %instant 
time $t$ of IF is replaced by LF time $ \tau = x^+ :=  ct+z$, where $z$ can be chosen as any one of the spatial coordinates.
Measurements at fixed LF time correspond to observations at the front of a light wave moving in the $-z$ direction, as in a flash photograph.
The LF %parametrization 
foliation of spacetime in terms of $(\tau, x^-)$, where $ x^- :=  ct-z$%is conjugate to $\tau$
, is distinct from $(t,z)$; however, the other spatial coordinates $(x,y)  := \bf{x_\bot}$ are the same in LF and IF.
Most important, the relation between $(\tau,x^-)$ and $(t,z)$ implies that physical descriptions in IF and LF are \emph{not} related to each other via a Lorentz transformation.

The $7$ kinematical LF generators correspond to invariances under spatial translations ($p^+:= p^0+p^z$, $\bf{p_\perp}$),
rotation about $\bf{z}$ ($J^z$),
and Lorentz boosts in the $\bf x$, $\bf y$ or $\bf x^+$ directions ($K_x$, $K_y$, $K_{x^+}$).
One of the $3$ dynamical generators is the LF Hamiltonian: the LF time evolution operator is $p^- \equiv p^0-p^z = i {d\over d\tau}$ and the eigenstates are eigenfunctions of the LF Hamiltonian, $\mathcal H_{LF} \vert \Psi \rangle = M^2 \vert \Psi \rangle $, $\mathcal H_{LF} = p^- p^+ -\bf{p}_\perp^2$.
The other two dynamical generators are the rotations about $\bf x$ or $\bf y$ ($J^x$, $J^y$). 
%The $2j+1$ mass-degenerate eigenstates: $- j, -j+1, \cdots J^z \cdots j-1, j$ determine the spin $j$ of the eigensolutions so the dynamical spin operators are not needed.
LF not only maximizes the number of kinematic generators of the Poincar\'e algebra, it also utilizes those which are typically most useful.
These features allow a complete, and in practice, Poincar\'e-invariant description of the system and its evolution. %in LF time $\tau$. 
In contrast, dynamical operators (beside the Hamiltonian) are necessary in IF; and since they describe spatial symmetries of the system, problematic pseudoforces must be included for a complete analysis.

A hadron, e.g. a proton, is an eigenstate of the QCD LF Hamiltonian: $\mathcal H_{LF} |\Psi_p\rangle = M_p^2 |\Psi_p\rangle$.
%\begin{equation}
%\label{LFSE}
%{\cal H}_{LF} |\Psi_p\rangle = M_p^2 |\Psi_p\rangle\,.
%\end{equation}
The  LF wave functions (LFWFs) for each Fock state of the proton, $\psi_p^{n_0}(x_i, \bf p_{\perp i}, \lambda_i)$, are projections $\langle n_0|\Psi_p\rangle$ of the eigenstate, where $\lambda_i$ are the LF helicity eigenvalues, which are kinematical quantities, and the $|n_0\rangle $ are the quark and gluon Fock state eigenstates of the free LF Hamiltonian $\mathcal H^0_{LF}$, which have the same quantum numbers as the proton.
One can identify the free (kinetic) and interaction-dependent contributions to the LF Hamiltonian
\begin{equation}
{\cal H}_{LF}  |\Psi_p\rangle = \mathcal H^0_{LF}  |\Psi_p\rangle + {\cal H}^{\rm int}_{LF}  |\Psi_p\rangle = M_p^2 |\Psi_p\rangle.
\end{equation}
The interactions, expressed in ${\cal H}^{\rm int}_{LF} $, couple the Fock states to each other:
\begin{equation}
\langle n_0|\mathcal H^0_{LF}  |\Psi_p\rangle  + \sum_{n_0^\prime} \langle n_0|\mathcal H^{\rm int}_{LF}|n_0^\prime\rangle \langle n_0^\prime  |\Psi_p \rangle = M_p^2 \langle n_0|\Psi_p\rangle \,.
\end{equation}
Consequently, each $\psi_P^{n_0}(x_i, \bf p_{\perp i}, \lambda_i)$ is a nontrivial Fock state LFWF, which is dynamically coupled to all the others.
Crucially, the wave functions of the eigenstates of $\mathcal H_{LF}$ are independent of the hadron's LF momenta $(p^+,\bf{p_\perp})$, \emph{viz}.\ they are independent of the frame defined by these quantities.
This property is manifest, e.g., in LF scattering formulae, such as the Drell-Yan--West expression for a form factor \cite{Brodsky:1997de}.
The LFWFs of any composite system are similarly frame-independent, in contrast to the analogous IF wave functions.
Thus the LF provides a fundamental, objective description of both elementary and bound-state systems without makeshift, frame-dependent dynamical corrections.

For instance, using LF, hadron form factors  can rigorously be computed from the overlap integral of the product of the initial- and final-state LFWFs \cite{Drell:1969km, West:1970av, Brodsky:1980zm}, see Fig.~\ref{Flo:Form-Factor in the IF}. The simplifying frame $q^+=0$ for the %virtual 
photon momentum $q$ eliminates %LF 
time-ordered contributions where the photon creates or annihilates quark-antiquark pairs.
Thus, %crucially, 
unlike IF, vacuum-induced currents are not generated. 
%by $q \bar q$ creation or annihilation from the vacuum when matrix elements of operators %such as ${\mathcal J}^\mu$ 
%are computed in the LF.
Note, however, that if one evaluates the electromagnetic current in a frame with $q^+ \ne 0$; one must allow for LF Fock states with extra $q \bar q$ pairs.

Factorization theorems, like the Drell-Yan formula \cite{Drell:1970wh} for lepton production $ p \bar p \to \ell \bar\ell X$, or the DGLAP \cite{Dokshitzer:1977sg, Gribov:1971zn, Lipatov:1974qm, Altarelli:1977zs} and ERBL~\cite{Lepage:1979zb, Efremov:1979qk} evolution equations are all derived using the LF,
which also enables a straightforward proof \cite{Brodsky:2000ii} that the anomalous gravitomagnetic moment of composite systems must vanish \cite{Okun:1962}.

%%%%%%%%%%%%%%%%%%%%%%%%%%%%%%%%%%%%%%%

\section{The GDH relation}
\label{SecGDH}

\begin{figure}[t]
\centering
\vspace{-0.cm}
 \captionsetup{width=0.89\linewidth}
\includegraphics[width=0.89\textwidth]{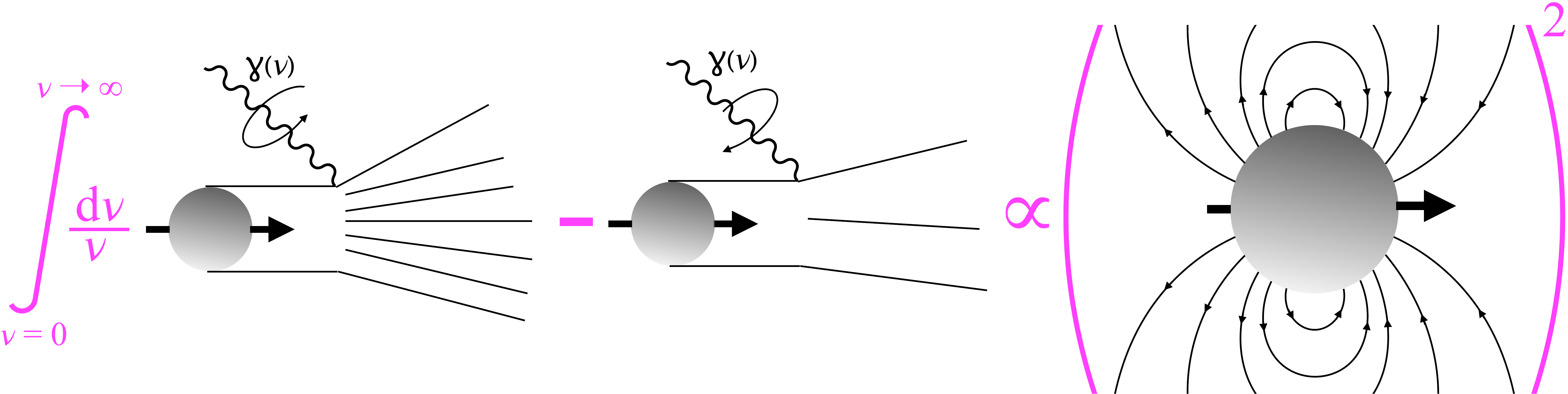}
\vspace{-0.15cm}
\caption{\label{Fig:GDH} 
{\bf The GDH relation.}
Polarized photons (wavy lines) impinging on a target particle (grey sphere) can produce secondary particles (thin lines). The flux of particles produced depends on the polarization direction of the photon (circular arrow) with respect to the target spin direction (straight arrow). The GDH relation connects the square of anomalous magnetic moment of the target particle, symbolized here by magnetic field lines, to the difference in the flux of produced particles for the two opposite photon polarizations. The relation holds after integrating the production difference over the photon energy $\nu$ (mathematical symbols in magenta).
}
\end{figure}

The GDH relation \cite{Gerasimov:1965et, Drell:1966jv, Hosoda:1966}, Fig. \ref{Fig:GDH}, %provides a remarkable connection of 
connects
the anomalous magnetic moment, $\kappa$, of a particle, such as a proton, to its helicity-dependent photo-production cross-sections, $ \sigma_P$, $ \sigma_A$:
\vspace{-0.2cm}
\begin{equation}
%\int_{\nu_0}^{\infty}{{\sigma_{P}(\nu) - \sigma_{A}(\nu)}\over{\nu}}d\nu={{4\pi^2 S \alpha\kappa^2}\over{M^2}}.
    \int_0         ^{\infty}{{\sigma_{P}(\nu) - \sigma_{A}(\nu)}\over{\nu}}d\nu={{4\pi^2 S \alpha\kappa^2}\over{M^2}}.
\label{eq:gdh}
\vspace{-0.2cm}
\end{equation}
Here,
$\nu= \sfrac{(k +p)^2}{2M}$ is the invariant energy of the absorbed photon of momentum $k^\mu$,
with $p^\mu$ the momentum of the studied particle,
$M$ is its rest mass,
$S$ is its spin,
and $\alpha$ is the QED fine-structure constant.
In $ \sigma_P$ ($ \sigma_A$), the photon spin is parallel (antiparallel) to the particle's spin.
%
%The relation %applies to all particles, e.g. an hadron, nucleus or electron, and
%has been empirically verified to within 10\% \cite{Helbing:2006zp}.
%
This relation, which is a fundamental feature of QFT,
can be used to illustrate how pseudoforces arise in IF from the dynamical character of IF boosts.
If pseudoforces are neglected, an incomplete description follows \cite{Barton:1967at}.
When they are incorporated \cite{Brodsky:1968xc, Brodsky:1968ea}, one arrives at the correct description in IF.

\subsection{Pseudoforces in the IF derivation.}
\label{Fictitious forces appearing}
The GDH relation has been derived in several ways:
(A) using a {\it dispersion relation} for the forward Compton amplitude \cite{Gerasimov:1965et, Drell:1966jv} together with the {\it low energy theorem} \cite{Low:1954kd, GellMann:1954kc}, the latter resting on Lorentz and gauge invariances;
(B) using LF current algebra \cite{Dicus:1972vp};
and
(C) using IF current algebra \cite{Hosoda:1966}.
Lorentz invariance is respected in (A) and (B), but not in (C) where a Lorentz-violating term appears on the right-hand side of Eq.~(\ref{eq:gdh}) \cite{Pradhan:1972ea, Pantforder:1997ii}:
%
%\vspace{-0.2cm}
%\begin{equation}
%    \int_ 0        ^{\infty}{{\sigma_{P}(\nu,M^2\nu^2/p^0) - \sigma_{A}(\nu,M^2\nu^2/p^0)}\over{\nu}}d\nu =
%4\pi^2 S \alpha \bigg({{ \kappa^2}\over {M^2}} - {{ (Z+\kappa)^2}\over {(p^0)^2}}\bigg),
%\label{eq:gdh-IF}
%\vspace{-0.2cm}
%\end{equation}
%
$4\pi^2 S \alpha \big({{ \kappa^2}\over {M^2}} - {{ (Z+\kappa)^2}\over {(p^0)^2}}\big),$
where $p^0$ is the initial state particle energy and $Z$ its %electric
charge.
The $p^0$ dependence violates Lorentz invariance, 
revealing $(Z + \kappa)^2$ as a fictitious elastic contribution to Compton scattering.

\subsection{Pseudoforces in the GDH relation for composite systems.}
\label{GDH deuteron}
Another example of pseudodynamics illustrated with the GDH relation appears when one explicitly accounts for particle compositeness.

The GDH relation was originally thought to be invalid for composite particles \cite{Barton:1967at}.
However, the purported invalidity originated from neglect of IF pseudoforces. % arising from the boost of IF wave functions.
%This error is analogous to overlooking inertial forces when working in a non-inertial frame.
%
The proof of Eq.~(\ref{eq:gdh}) for the deuteron was given in Refs.\,\cite{Brodsky:1968xc, Brodsky:1968ea}.
Verifying Eq.~(\ref{eq:gdh}) for the deuteron involves the analysis of a bound-state of two nucleons in an external electromagnetic field.
A boost is required because the deuteron momentum is affected by the probing photon, integrated in Eq.~(\ref{eq:gdh}) over all values of $\nu$.
The covariant Bethe-Salpeter formalism \cite{Salpeter:1951sz}
provides
%can be used to derive the Foldy-Wouthuysen \cite{Foldy:1949wa} transformation and thereby obtain 
the correct IF boost of the deuteron wave function, which includes pseudodynamics.
In contrast, assuming a product of boosts of two independent nucleons \cite{Barton:1967at} overlooks pseudodynamics and produces an erroneous IF result.

The boosted IF deuteron wave function includes Dirac spinors of the form \cite{Brodsky:1968xc, Brodsky:1968ea}:
\begin{equation}
\biggl[ \begin{matrix} 
1+ {{\bm{\sigma_i  \cdot P}}\over{\mathcal{M}+E}} {{\bm{\sigma_i  \cdot p}}\over{2m_i + k_i}} \\
\bm{\sigma_i  \cdot} \big( {{\bm{P}\over{\mathcal{M}+E}} + {{\bm{p}}\over{2m_i + k_i}} \big)}  
\end{matrix} \biggr]
\end{equation}
where:
$\bm{\sigma_i}$, $i=a,b$, is the spin of nucleon $i$;
$(E,\bm{P})=p_a+p_b$ is the deuteron 4-momentum, with $p_i$ the center-of-mass nucleon 4-momenta;
$\bm{p}=(m_a \bm{p_a} - m_b \bm{p_b})/\mathcal{M}$ is the relative momentum;
$\mathcal{M}$ is the deuteron rest mass;
$m_{1,2}$ the nucleon masses;
and $k_i \approx \bm{p}^2/(2m_i)$;

The boost generates the $\bm{\sigma} \cdot (\bm{p},\bm{P})$ terms.
They would be absent if one erroneously assumes that the net boost is mrely the product of boosts for two independent nucleons.
They arise from the boost-induced mixing of the triplet state $\ket {1}$ of the deuteron $\ket {1/2} \times \ket{1/2}$ system with its singlet state $\ket {0}$~\cite{McGee:1967zza}.
Thus a pseudointeraction appears in the resulting boosted Pauli equation for a composite system in addition to the sum of the individual one-body interactions with the external field.
The extra ``spin-orbit'' terms express the spin coupling to the induced magnetic field $\bm{B}_{\rm ind} = \bm{v} \times \bm{E}$ owing to motion in the external electric field.

The pseudoterms $\bm{\sigma} \cdot (\bm{p},\bm{P})$ indicate that energy from the external field is changed into deuteron internal energy.
Specifically, the force acting on the two-body system is modified during a boost owing to triplet-singlet state mixing, since only the triplet state 
$\ket {1}$ couples to the external magnetic field.
Thus, a kinematic action -- the IF boost -- results in a dynamical modification:
the change of force, from the reduced coupling to the external magnetic field owing to the transfer of  spin-1 $\to$ spin-0.
Without this effect, the GDH relation, as dynamically determined in IF, cannot be recovered for composite objects.
The kinematical and dynamical mixing is analogous to classical mechanics, where pseudoforces mix kinetic and potential energies, e.g., when the kinetic energy is converted into the energy of a centrifugal force if a rotating non-Galilean frame is chosen for the analysis.

We have discussed Eq.\,(\ref{eq:gdh}) within the IF framework.
It can be derived for composite systems using the frame-independent LF formalism, and  pseudoforces do not appear.
This offers a concrete example of the advantage of LF.
Although IF derivation of the GDH relation is possible -- with considerably more challenges than in LF, as illustrated above -- this is not so for many other observables.

%%%%%%%%%%%%%%%%%%%%%%%%%%%%%%%%%%%%%%%

\section{Lorentz Contraction}
\label{L.C.}
%
%Here, we discuss  whether Lorentz contraction should be considered a fictitious effect.
Lorentz contraction is clearly a subjective effect because it depends explicitly on the observer's motion relative to the contracted object; but is it fictitious?
Lorentz contraction is often invoked when describing relativistic collisions of composite objects.
However, mistakenly ascribing an objective meaning to Lorentz contraction can lead to erroneous descriptions.
In fact, since Lorentz contraction is not a true observable, it cannot influence any observable properties of a relativistic collision.
Working with the LF approach enables one to avoid such pitfalls.

In discussions of collisions involving %one or more 
composite objects moving with relativistic velocities, one commonly depicts the colliding objects as thin discs (pancakes), in an attempt to highlight Lorentz contraction \cite{Wolschin:2020qxa}.
However, if such contractions were fundamental to the dynamical description of the collision, a paradox would ensue because the internal dynamics of hadrons is frame independent.
For instance, the structure of the proton measured in deep inelastic lepton scattering %$e p \to e^\prime X$ 
in the laboratory frame, where the proton is at rest, is the same~\cite{ParticleDataGroup:2020ssz} as that when measured at a lepton-proton collider, where the proton is% relativistic; hence
, supposedly, Lorentz-contracted.

This seeming conflict between the Lorentz contraction of a moving object versus the frame-independence of hadron observables is resolved by carefully distinguishing the character of the measurement \cite{Terrell:1959zz, Penrose:1959vz, Weisskopf}.
The length of an object is defined by two observations performed at identical Galilean time, $t$.
Thus, in addition to violating causality, the Lorentz contraction would only be applicable for an idealized experiment in which a prodigious number of detectors is placed close to the object and arranged to image it at %a given 
fixed $t$.
However, in actual measurements, such as lepton-proton scattering, the exchanged photon interacts with the quarks of a proton at fixed LF time, $\tau$, i.e., along the front of a light-wave, in analogy with a flash photograph, \underline{not at fixed $t$}.
One also measures at fixed $\tau$ when observing radiation from a %composite 
system at a detector far from the moving source.
That no Lorentz contraction is observed in actual measurements is immediately clear when using LF, but it requires non-trivial reasoning when using IF.
The often-used illustration of relativistic nuclei as colliding ``pancakes"  is therefore not only subjective, but also erroneous.

%%%%%%%%%%%%%%%%%%%%%%%%%%%%%%%%%%%%%%%

\section{Vacuum energy}
 \label{vacuum}
Estimating the ``vacuum energy'', a conjectured source for the cosmological constant, $\Lambda$,
has been a notorious problem: conventional arguments yield values up to $10^{120}$-times greater than observation \cite{Zee:1983jg, Weinberg:1988cp, Carroll:2000fy, Peebles:2002gy, Padmanabhan:2002ji, Zee:2008zza}.
However, such analyses rely on the concept of IF vacuum fluctuations, which does not provide objective,  frame-independent, properties of spacetime.
In contrast, LF dynamics predicts $\Lambda=0$ \cite{Brodsky:2009zd, Brodsky:2010xf, Chang:2011mu, Brodsky:2012ku, Cloet:2013jya} because the LF vacuum does not fluctuate:
the vacuum has $p^+ = 0$, by definition; 
momentum conservation demands that some virtual particles participating in fluctuations would have $p^+<0$, in conflict with the LF property that $p^+>0$ for any particles.
% $p^+ = E + p^z$;
%($E$ is the system energy, and $p^z$ its 3-momentum along $z$);
%the vacuum has $p^+ = 0$ by definition;
%momentum conservation demands that some virtual particles participating in fluctuations have $p^+<0$;
%and this is forbidden by LF dynamics.
%
%The $\Lambda=0$ LF prediction agrees with the fact that cosmological observations \cite{Riess:1998cb, Perlmutter:1998np}, which suggest a small $\Lambda \neq 0$, may instead be understood with $\Lambda = 0$ if one properly accounts for QCD-like aspects of general relativity \cite{Deur:2017aas}.

Just as pseudodynamics emerge to preserve fundamental principles, in classical mechanics and in IF dynamics, IF vacuum fluctuations arise to preserve causality, a keystone of QFT.
To see this explicitly, one may trace how IF vacuum fluctuations arise; this will also provide an intuitive understanding of why the LF vacuum is trivial, complementing the formal reason that $p^+>0$ for all particles.

\subsection{The origin of the IF vacuum fluctuations.}
\begin{figure}[t]
\centering
\vspace{-0.cm}
 \captionsetup{width=0.89\linewidth}
 \begin{tabular}{ccc}
\includegraphics[clip, width=0.6\textwidth]{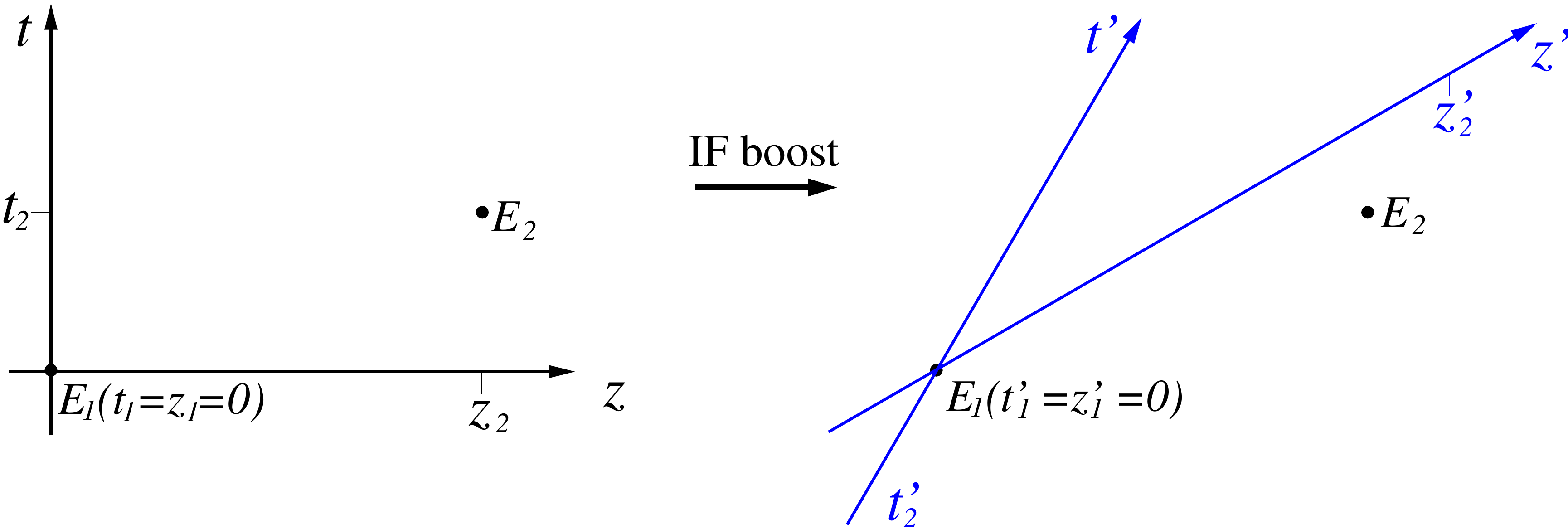} & \rule{3em}{0ex} &
\hspace{-1.5cm}
\includegraphics[clip, width=0.32\textwidth]{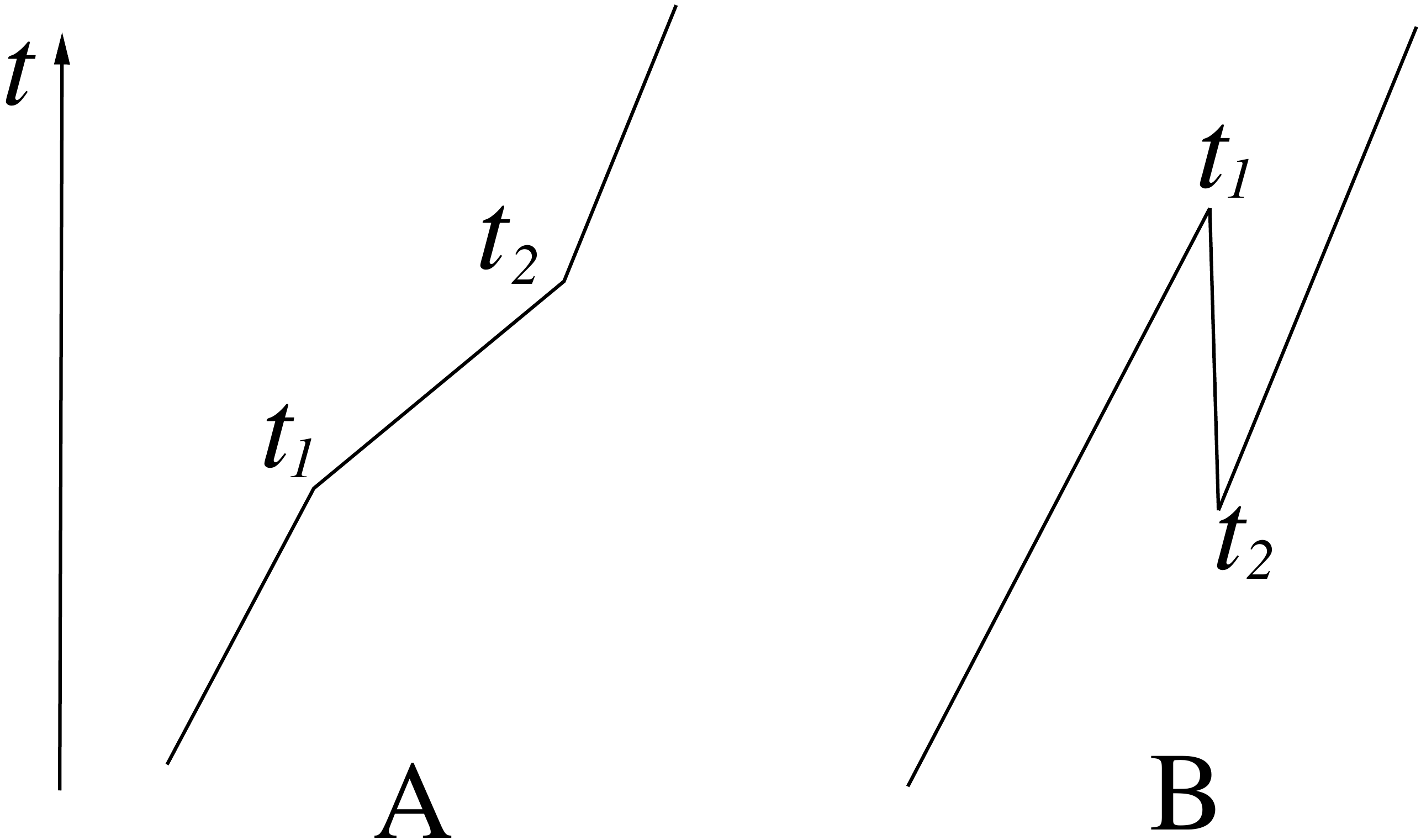}
\end{tabular}
\vspace{-0.1cm}
\caption{\label{Path-Int-NLO-prop}
{\bf Reordering of event times by an IF boost.}
{\it Left panel}. IF spacelike propagation from events $E_1$ to $E_2$, in frame $(t,z)$ (with $t_2-t_1 > 0$). {\it Middle panel}. Same, but in a frame $(t',z')$ (with $t'_2-t'_1 < 0)$, boosted with respect to $(t,z)$: since an IF boost in the $z$-direction rotates clockwise the $t$-axis and counterclockwise the $z$-axis by the same angle, $E_2$ may occur before $E_1$ in the boosted $(t',z')$ frame: $t'_2<t'_1$.
%Event $E_1$ is the origin of the systems.
%
{\it Right panel}.
A) Single particle propagation in frame $(t,z)$.  B) Same process in frame $(t',z')$ (pair creation).
%In contrast, LF boosts only rescale the axes, thus always preserving events ordering.
}
\end{figure}
IF vacuum fluctuations arise when one allows propagators to link spacelike events.
This is explained in Ref.\,\cite{Feynman:1987gs}, which considers a scalar particle, $\phi$, interacting with an external scalar potential, $U$.
The amplitude describing the state survival probability after scattering events $ E_1(x_1)$ and $E_2(x_2)$ is $\mathcal{A}=1-{\mathcal I}$, 
\begin{equation}
{\mathcal I}=\int dx_1 dx_2\int {{dP}\over{(2\pi)^3 4E^2}}
U^*(x_2)\phi^*( x_2)e^{-i P_\mu (x_2^\mu - x_1^\mu) }U( x_1)\phi( x_1),
\label{if_prop}
\end{equation}
with $P$ the four-momentum of particle $\phi$, of positive energy.
${\mathcal I}$ can be nonzero for spacelike $ x_2^\mu -  x_1^\mu$ because any function, $f(t)$, that can be Fourier decomposed using only positive energies:
%only, i.e.,
%
\begin{equation}
f(t)=\int_0^\infty dE e^{-iEt} F(E),
\label{Fourier}
\end{equation}
cannot be zero for a finite $t$ range (unless it vanishes everywhere); so, spacelike processes are allowed.

For such processes, one may have $t_2>t_1$ in some frames, such as the frame $(t,z)$ of Fig.~\ref{Path-Int-NLO-prop}\,--\,left.
Here, ${\mathcal I}$ describes the propagation of a single particle, see image~A in Fig.~\ref{Path-Int-NLO-prop}\,--\,right.
In another frame %, the same process may have 
$t_2<t_1$, as for frame $(t',z')$ in Fig.~\ref{Path-Int-NLO-prop}\,--\,center, where $\mathcal{I}$ describes an initially propagating particle, which then annihilates at $t_1$ with an antiparticle from a pair created at $t_2$; see image~B in Fig.~\ref{Path-Int-NLO-prop}\,--\,right.
However, additional amplitudes from disconnected vacuum diagrams must then be added to preserve unitarity: the vacuum diagrams compensate the negative amplitudes of the Z-graphs of image B.
Disconnected vacuum processes are thus directly coupled to the causality-violating processes imposed by Eq.~(\ref{Fourier}).
In contrast, in LF dynamics, a boost simply rescales the $\tau$ and $x^-$ axes and thus preserves the sign of the LF time interval $\tau_2-\tau_1$.
Therefore causality is preserved, no Z-graphs appear; hence, disconnected vacuum  processes are not required to enforce unitarity.

\begin{figure}[t]
\centering
\vspace{-0.cm}
 \captionsetup{width=0.89\linewidth}
\includegraphics[width=16.0cm]{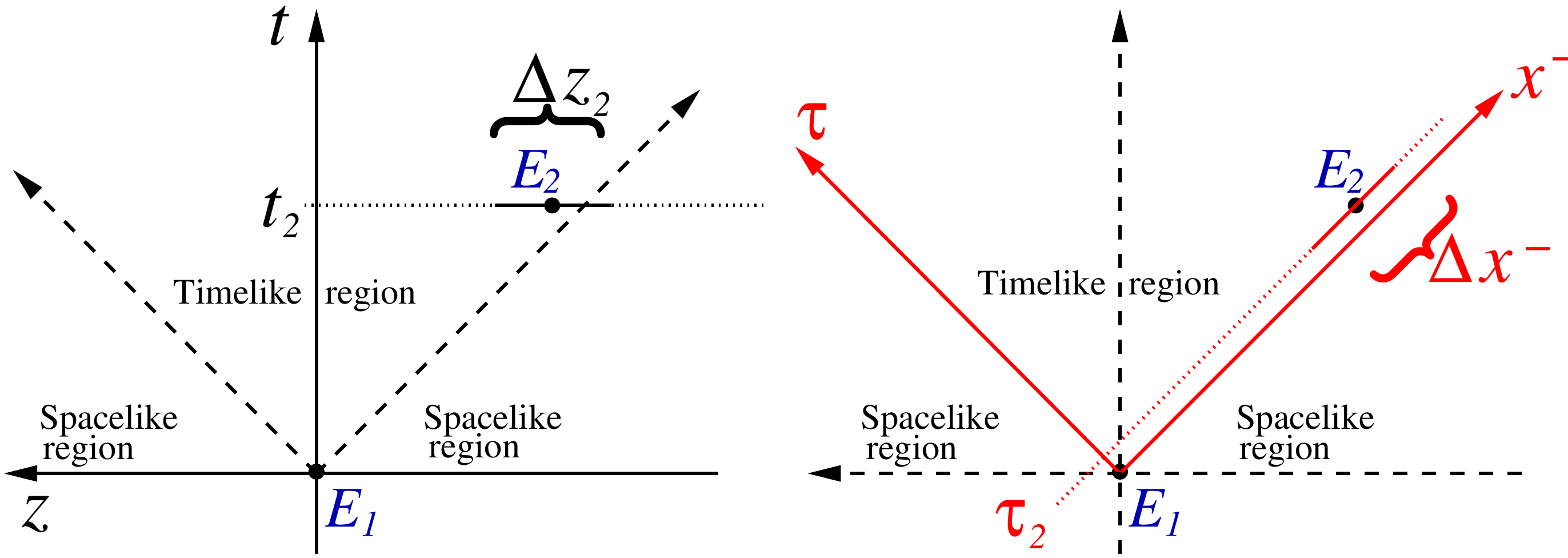}
%\vspace{-0.05cm}
\caption{\label{Eisenberg} 
{\bf Effect of quantum fluctuations on the causal nature of an event.}
Events in the spacelike region with respect to event $E_1$ generate vacuum fluctuations at the origin of the vacuum energy.  
Left panel: In IF dynamics, the Heisenberg uncertainty $\Delta z_2$ (horizontal segment around event $E_2$) applies to fixed time $t_2$. Thus, $\Delta z_2$ may allow $E_2$ (which would otherwise be timelike) to be spacelike.
Right panel: In LF dynamics, the  uncertainty $\Delta x^-$ (red segment around $E_2$) is at fixed LF time $\tau$, i.e. along the border between the timelike and spacelike regions. Therefore, $E_2$ remains timelike with respect to $E_1$, and vacuum fluctuations are not possible.
}
\end{figure}

Since the integrals in Eqs.\,(\ref{if_prop}), (\ref{Fourier}) are suppressed for large values of $  P_\mu(  x_2^\mu -   x_1^\mu)$, this discussion connects to Heisenberg's principle, often used to
depict the origin of vacuum fluctuations:
Fig.~\ref{Eisenberg} illustrates how the $\Delta P_z \Delta z$ uncertainty for $E_2$, in-principle
timelike, may actually allow it to be spacelike, enabling disconnected vacuum processes. In contrast, in LF, $\Delta x^-$ varies along the light-front direction $x^-$, preserving the timelike nature of $E_2$.

%Mathematics which violate basic principles (in this case, causality) need not be accepted.
The violation of causality in Eq.~(\ref{Fourier}) can be avoided by quantizing the field at fixed $\tau$ (LF dynamics), so that two causally linked events cannot become spacelike owing to quantum fluctuations, see Fig.~\ref{Eisenberg}.
Thus the initial mechanism which leads to vacuum fluctuations, via unitarity, is eliminated: the LF vacuum is trivial, and LF dynamics preserves causality.

Clearly, the complex IF vacuum utilizes a fictitious mechanism for restoring causality, which in any case is broken by choosing a dynamics that violates Poincar\'e invariance.

\subsection{Confinement contains condensates.}
Insofar as contributions to $\Lambda$ from vacuum fluctuations are concerned, the answer might be straightforward \cite{Brodsky:2012ku}.
The connection between vacuum structure and $\Lambda$ is typically linked to a belief that infinitely many distinct, spacetime-independent condensates fill the Universe.
These are the supposedly measurable consequences of IF vacuum fluctuations.
However, the problem vanishes if one jettisons the notion that condensates have a physical existence, independent of hadronic matter \cite{Brodsky:2009zd}.
This seems obligatory in a confining theory \cite{Chang:2011mu, Brodsky:2010xf, Brodsky:2012ku, Cloet:2013jya} such as QCD where all  %measurable consequences of the theory, including its ground state, 
predictions 
can be expressed using a hadronic basis.
This is {\it quark-hadron duality}.
The paradigm drawn in Refs.\,\cite{Brodsky:2009zd, Chang:2011mu, Brodsky:2010xf, Brodsky:2012ku, Cloet:2013jya} can therefore be summarized as follows: ``If quark-hadron duality is a reality, then condensates, those quantities %which have been 
commonly viewed as constant mass-scales that fill all spacetime, are instead wholly contained within hadrons; to wit, they are properties of hadrons themselves, expressed, e.g., in their LFWFs.''

The chiral condensate, $\langle \bar q q \rangle$, is the vacuum condensate most often discussed.
It is an order parameter for dynamical chiral symmetry breaking and, using LF quantization, it was long ago argued \cite{Casher:1974xd} to be a property of hadron wave functions, not the vacuum.
This quality is not surprising: in a rigorous treatment of phase transitions in statistical mechanics, a transition only occurs in the infinite-volume limit.
However, in empirical reality, e.g., %magnetization or 
Cooper pair condensation, the order parameters are only observed in finite-volume samples; {\it viz}.\ the condensates are found within the material that supports them, not throughout an infinite volume.
Similarly, especially because of confinement, QCD condensates should be contained within the same domain which permits the propagation of the gluons and quarks that produce them; namely, inside hadrons.
%Importantly, that condensates are contained within hadrons agrees with all empirical observations \cite{Brodsky:2012ku}.
%
Recognizing that QCD condensates are ``in-hadron" properties, 
reduces the mismatch between observation and theory by a factor of $10^{46}$.

%%%%%%%%%%%%%%%%%%%%%%%%%%%%%%%%%%%%%%%

\section{Conclusion}
In classical dynamics, one may choose any frame to analyze a system -- provided that pseudoforces are included if the frame is non-inertial.
Likewise, one may %in principle 
choose the IF approach, if its advantages balance the complications of
including pseudoeffects.
An example of IF dynamics, supplemented by covariant methods to compute pseudocontributions, is the GDH relation derivation  for the deuteron \cite{Brodsky:1968xc, Brodsky:1968ea}.
Yet, this is not a fundamental derivation because it not only demands additional input beyond those intrinsic to QED and the nuclear force framework, but also introduces subjective, observer-dependent forces.
Of greater practical concern in the case of systems described by  %nonperturbative  QCD, such as hadronic matter, 
QCD, the treatment of the pseudoeffects in the IF is generally intractable.
The inability  to provide a frame-independent derivation of the GDH relation, even for non-composite systems;
and the erroneous suggestion that Lorentz contraction plays a dynamical role in relativistic heavy-ion collisions
are other examples of the drawbacks of IF quantization.
The contrast between the complex, causality-violating IF vacuum and the simple LF vacuum demonstrates the usefulness of LF quantization and emphasizes that the cosmological constant cannot receive a physical contribution from vacuum structure.

Today, no approach to QFT is perfect, but LF quantization possesses many merits.  It often provides the simplest known description of Nature.  
%This is, perhaps, most true in the proofs of exact results for hard QCD processes, or of the vanishing of the gravitomagnetic moment of composite systems.
%
As always in physics, one can formulate a problem using any framework one desires; but if the wrong approach is chosen, the costs are high \cite{Weinberg:1981qq}.

It is worth closing with a lesson from classical mechanics.
Although one may elect to describe the solar system using a geocentric standpoint, perhaps because it appears intuitive, one will eventually learn that this is not the minimal description.
Ultimately, the heliocentric vision is more economical and objective; and arguably, closer to Nature.
Likewise, despite its initial conflict with human-based intuition, the absence of subjective effects in LF quantization often makes it the most economical choice for a description of relativistic dynamics.

\begin{addendum}
 \item [Acknowledgments] This work is supported by:
U.S.\ Department of Energy contract DE-AC02-76SF00515 (SJB);
%SLAC-PUB-17618
U.S.\ Department of Energy, Office of Science, Office of Nuclear Physics, contract DE-AC05-06OR23177 (AD);
and National Natural Science Foundation of China, grant no.\,12135007 (CDR).

\item[Author Contributions] All authors contributed equally to the preparation of this contribution.

\item[Competing Interests] The authors declare that they have no competing financial interests.
\end{addendum}

\noindent{\bfseries References}\setlength{\parskip}{12pt}%

%%\bibliographystyle{naturemag}
%%\bibliography{BiBFictitious}

\end{document}